\documentclass[twocolumn,superscriptaddress,prb]{revtex4}
\usepackage{amsmath}
\usepackage{graphicx}
\usepackage{xcolor} 
\usepackage[normalem]{ulem} 

\newcommand{\stkout}[1]{\ifmmode\text{\sout{\ensuremath{#1}}}\else\sout{#1}\fi}
\newcommand{\md}{\mathrm d}

\begin{document}

\title{Efficient two-dimensional control of barrier crossing}

\author{Steven Blaber}
\email{sblaber@sfu.ca}
\affiliation{Dept.~of Physics, Simon Fraser University, Burnaby, British Columbia V5A 1S6, Canada}
\author{David A.\ Sivak}
\email{dsivak@sfu.ca}
\affiliation{Dept.~of Physics, Simon Fraser University, Burnaby, British Columbia V5A 1S6, Canada}

\begin{abstract}
Driven barrier crossings are pervasive in optical-trapping experiments and steered molecular-dynamics simulations. Despite the high fidelity of control, the freedom in the choice of driving protocol is rarely exploited to improve efficiency. We design protocols that reduce dissipation for rapidly driven barrier crossing under two-dimensional control of a harmonic trapping potential, controlling both trap center and stiffness. For fast driving, the minimum-dissipation protocol jumps halfway between the control-parameter endpoints. For slow driving, the minimum-dissipation protocol generically slows down and tightens the trap as it crosses the barrier, resulting in both significant energy savings and increased flux compared to naive and one-dimensional protocols (that only change trap center). Combining fast and slow results, we design protocols that improve performance at all speeds.
\end{abstract}

\maketitle

{\it Introduction.}---Modern advances in single-molecule biophysics make possible the precise spatial and temporal control of biological systems. Optical tweezers can be used to probe the conformational and energetic properties of biopolymers (DNA and RNA molecules)~\cite{liphardt2002,collin2005,bustamante2000,bustamante2003,woodside2006,neupane2017} and molecular machines (ATP synthase~\cite{Toyabe2011,toyabe2012,kawaguchi2014}, kinesin~\cite{svoboda1993,svoboda1994,kojima1997,hunt1994}, and myosin~\cite{greenberg2016,Laakso2008,norstrom2010,nagy2013}). Additionally, computer simulations such as steered molecular dynamics have the freedom to fully control the molecular and trapping potentials. Despite the relative freedom of control, experiments and simulations rarely exploit the possibility of optimized control protocols, and the few that do are generally limited to optimization of a single control parameter~\cite{Dellago2014,Geiger2010,rico2021}. In this letter, we design minimum-dissipation protocols for harmonic trapping potentials under two-dimensional control of trap center and stiffness, allowing for the specification of the time-dependent mean and variance and resulting in significantly reduced dissipation compared to control over the trap center only. We find generic features that can be readily applied to biophysical experiments and simulations.

We are interested in describing micro- and nano-scale thermodynamic systems. Due to the small scale, thermal fluctuations play a significant role in the dynamics, and the systems are best described by stochastic thermodynamics. Stochastic thermodynamics typically describes the nonequilibrium transformation of heat, work, and entropy of small-scale fluctuating systems~\cite{Jarzynski2011,Seifert2012}. This field has seen significant growth over the last couple decades, stemming from important results such as the Jarzynski equality~\cite{Jarzynski1997} and Crooks fluctuation theorem~\cite{Crooks1999}. 

One important implication of the Jarzynski equality is that it allows for the determination of equilibrium free-energy differences from nonequilibrium work measurements. However, the accuracy of the free-energy estimate decreases with increasing dissipation~\cite{Gore2003,Jarzynski2006}. Designing protocols that dissipate less energy can therefore improve the accuracy of free-energy estimates~\cite{Blaber2020Skewed}. Additionally, in order to maintain complex nonequilibrium order, molecular machines must operate rapidly, potentially incurring large energetic costs~\cite{Brown2019}. The dissipation incurred from rapid driving can be mitigated by designing less-dissipative protocols~\cite{OptimalPaths,deffner2020,bonancca2014}, resulting in improved performance. 

Linear-response theory has been used to derive a thermodynamic-geometry framework to guide design of single and multidimensional minimum-dissipation protocols in general stochastic-thermodynamic systems~\cite{OptimalPaths}. This framework has been applied to DNA-hairpin pulling experiments, demonstrating that one-dimensional designed pulling protocols significantly reduce dissipation~\cite{Tafoya2019}. The DNA-pulling process can be reasonably well described as a driven barrier crossing~\cite{neupane2015}, where the barrier to be overcome is the transition state between folded and unfolded conformational ensembles. One-dimensional minimum-dissipation control of driven barrier crossing slows down as the trap crosses the energy barrier~\cite{Sivak2016}.

\begin{figure}
	\includegraphics[width=\linewidth]{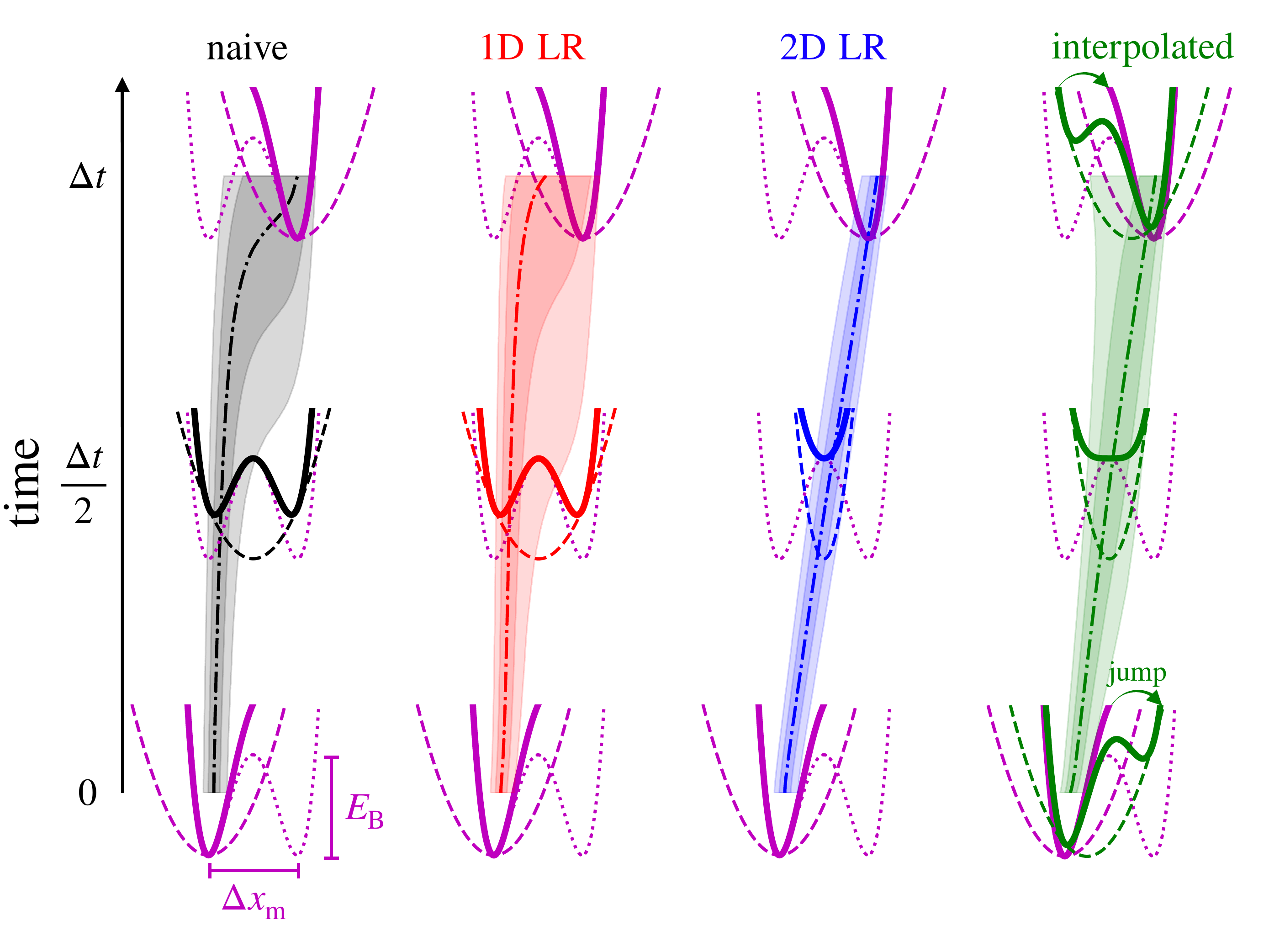}
	\caption{Time-dependent protocols for driven barrier crossing at intermediate protocol duration. Naive (black), one-dimensional linear response (1D LR, red), two-dimensional linear response (2D LR, blue), and interpolated (green). Snapshots of the total (solid), static hairpin (dotted), and time-dependent trap (dashed) potential are shown for $t=0$, $\Delta t/2$, and $\Delta t$. The hairpin, initial, and final potentials are the same across protocols (purple). Dash-dotted curves: median positions during corresponding protocol. Shading: 9\%, 25\%, 75\%, and 91\% quantiles, which are approximately evenly spaced for a Gaussian distribution. Barrier height is $E_{\rm B} = 4k_{\rm B} T$, initial and final trap stiffnesses are $k_{\rm i} = k_{\rm f} = 4 k_{\rm B}T/x_{\rm m}^2$, and protocol duration is $\tau_{\rm D}$ for diffusion time $\tau_{\rm D} = \Delta x_{\rm m}^2/(2D)$.} 
	\label{Protocol}
\end{figure}

\begin{figure*} 
	\includegraphics[width=\textwidth]{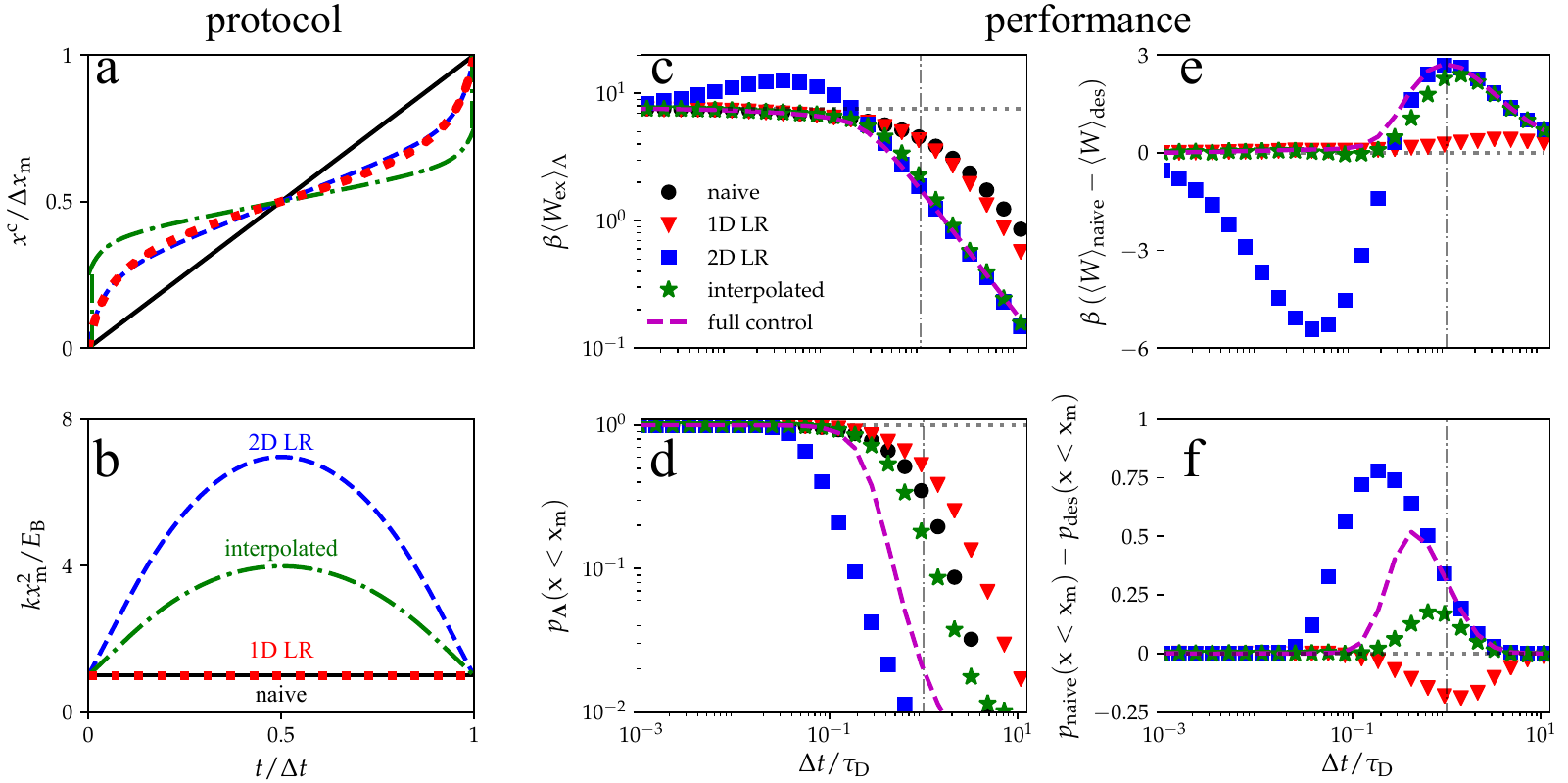} 
	\caption{Performance of the naive (black), 1D LR (red), 2D LR (blue), and interpolated (green) protocols. Protocols show (a) trap center and (b) trap stiffness, as a function of time $t$ normalized by protocol duration $\Delta t$. The interpolated protocol is shown for a duration $\Delta t = \tau_{\rm D}$. (c) Excess work $\langle W_{\rm ex} \rangle_{\boldsymbol{\Lambda}}$, (d) probability $p_{\boldsymbol{\Lambda}}(x<x_{\rm m})$ that a trajectory does not cross the barrier, (e) difference in work between designed and naive protocols, and (f) difference in $p_{\boldsymbol{\Lambda}}(x<x_{\rm m})$ between naive and designed protocols, all as functions of protocol duration $\Delta t/\tau_{\rm D}$ scaled by diffusion time $\tau_{\rm D}$. Purple dashed curve: optimal-transport process under full control.}
	\label{Work_distance}
\end{figure*}

We explore two-dimensional control (of both trap center and stiffness) of driven barrier crossing. This greater control allows specification of both the time-dependent mean and variance of the position distribution, and results in a qualitatively distinct designed protocol (Fig.~\ref{Protocol}). Such a designed protocol has jumps at the start and end that decrease in size as the duration increases, and slows down and tightens as it crosses the barrier, approximately linearly driving the mean and maintaining roughly constant variance throughout the protocol. For any duration, the designed protocols significantly improve performance in terms of both dissipation and flux compared to naive and one-dimensional control (Fig.~\ref{Work_distance}).

{\it Theory.}---Consider a system obeying Fokker-Planck dynamics,
\begin{align}
	\beta\gamma\frac{\partial p_{\boldsymbol{\Lambda}}(x,t)}{\partial t}  = -\frac{\partial }{\partial x}\left\{\beta f_{x}[x,\boldsymbol{\lambda}(t)]-\frac{\partial }{\partial x}\right\}p_{\boldsymbol{\Lambda}}(x,t) \ ,
	\label{Fokker Planck}
\end{align} 
describing the overdamped motion of a continuous degree of freedom $x$ with damping coefficient $\gamma$ driven by a time-dependent conservative force $f_{x}[x,\boldsymbol{\lambda}(t)] = -\partial V_{\rm tot}[x,\boldsymbol{\lambda}(t)]/\partial x$, for internal energy $V_{\rm tot}(x,\boldsymbol{\lambda})$. $p_{\boldsymbol{\Lambda}}(x,t)$ is the probability distribution over microstates $x$ at time $t$ given the control protocol $\boldsymbol{\Lambda}$. The system is in contact with a heat bath at temperature $T$ such that the equilibrium probability distribution over $x$ at fixed control parameters $\boldsymbol{\lambda}$  is $\pi(x|\boldsymbol{\lambda}) \equiv \exp\{\beta[F(\boldsymbol{\lambda})-V_{\rm tot}(x,\boldsymbol{\lambda})]\}$, for free energy $F(\boldsymbol{\lambda}) \equiv -k_{\rm B}T\, \ln\sum_{x}\exp[-\beta V_{\rm tot}(x,\boldsymbol{\lambda})] $, where $\beta \equiv (k_{\rm B}T)^{-1}$ for Boltzmann's constant $k_{\rm B}$. The average excess work $W_{\rm ex} \equiv W-\Delta F$ by an external agent changing control parameters $\boldsymbol{\lambda}$ according to protocol $\Lambda$ is
\begin{align}
\langle W_{\rm ex} \rangle_{\boldsymbol{\Lambda}} = -\int_{0}^{\Delta t }\md t \,\frac{\md \boldsymbol{\lambda}^{\rm T}}{\md t} \langle \delta {\bf f}_{\boldsymbol{\lambda}}(t) \rangle_{\boldsymbol{\Lambda}} \ ,
\label{work def}
\end{align}
where superscript ${\rm T}$ denotes transpose. ${\bf f}_{\boldsymbol{\lambda}} \equiv -\partial V_{\rm tot}/\partial\boldsymbol{\lambda}$ are the forces conjugate to the control parameters, and $\delta {\bf f}_{\boldsymbol{\lambda}} \equiv {\bf f}_{\boldsymbol{\lambda}} - \langle {\bf f}_{\boldsymbol{\lambda}} \rangle_{\boldsymbol{\lambda}}$ the deviations from the equilibrium averages. Angle brackets $\langle \cdots\rangle_{\boldsymbol{\Lambda}}$ denote a nonequilibrium ensemble average given the control protocol $\boldsymbol{\Lambda}$ and $\langle \cdots\rangle_{\boldsymbol{\lambda}}$ an equilibrium average given control-parameter values $\boldsymbol{\lambda}$.

{\it Slowly driven systems.}---In the quasistatic (infinitely slow) limit, the probability distribution remains at equilibrium throughout the protocol, and the excess work approaches zero. For long-but-finite protocol duration, linear-response (LR) theory yields the leading-order contribution to the excess work~\cite{OptimalPaths},
\begin{align}
\label{LR excess work}
\langle W_{\rm ex}\rangle_{\boldsymbol{\Lambda}} \approx \int_{0}^{\Delta t}\md t \ \frac{\md {\boldsymbol{\lambda}}^{\rm T}}{\md t} \ \zeta[\boldsymbol{\lambda}(t)] \ \frac{\md {\boldsymbol{\lambda}}}{\md t} \ ,
\end{align}
in terms of the generalized friction tensor with elements
\begin{align}
\zeta_{j \ell}(\lambda) \equiv \beta \int_0^{\infty} \md t \, \langle \delta f_{\lambda_{j}}(t) \delta f_{\lambda_{\ell}}(0)\rangle_{\boldsymbol{\lambda}} \ . 
\label{friction}
\end{align}
$\zeta_{j \ell}$ is the Hadamard product $\beta \langle\delta f_{\lambda_{j}} \delta f_{\lambda_{\ell}}\rangle_{\boldsymbol{\lambda}} \circ \tau_{j \ell}$ of the conjugate-force covariance (the force fluctuations) and the integral relaxation time
\begin{align}
\label{relax1}
\tau_{j \ell} \equiv \int_0^{\infty} \md t \, \frac{\langle \delta f_{\lambda_{j}}(t) \delta f_{\lambda_{\ell}}(0)\rangle_{\boldsymbol{\lambda}}}{\langle \delta f_{\lambda_{j}} \delta f_{\lambda_{\ell}}\rangle_{\boldsymbol{\lambda}}} \ ,
\end{align}
the characteristic time for these fluctuations to die out.

For overdamped dynamics, the friction can be calculated directly from the total energy as~\cite{zulkowski2015}
\begin{align}
\zeta_{j\ell}(\boldsymbol{\lambda}) = \int_{-\infty}^{\infty} \md x 
\, 
\frac{\partial_{\lambda_{j}}\Pi_{\rm eq}(x,\boldsymbol{\lambda})\partial_{\lambda_{\ell}}\Pi_{\rm eq}(x,\boldsymbol{\lambda})}{\pi_{\rm eq}(x,\boldsymbol{\lambda})}
\ ,
\label{CDF friction}
\end{align}
where $\Pi_{\rm eq}(x,\boldsymbol{\lambda}) \equiv \int_{-\infty}^{x}\md x'\pi_{\rm eq}(x',\boldsymbol{\lambda})$ is the equilibrium cumulative distribution function and $\partial_{\lambda_{j}}$ is the partial derivative with respect to $\lambda_{j}$.

Within the linear-response approximation, the excess work is minimized by a protocol with constant excess power~\cite{OptimalPaths}. For a single control parameter, this amounts to proceeding with velocity $\md \lambda^{\rm LR}/\md t \propto \zeta(\lambda)^{-1/2}$, which when normalized to complete the protocol in a fixed allotted time $\Delta t$, gives 
\begin{align}
\label{lambdaoptdot}
\frac{\md \lambda^{\rm LR} }{\md t}= \frac{\Delta \lambda}{\Delta t}\frac{\overline{{\zeta}^{1/2}}}{\sqrt{\zeta(\lambda)}} \ ,
\end{align}
where the overline denotes the spatial average over the naive (linear) path between the control-parameter endpoints.

For multidimensional control, the minimum-dissipation protocol solves the Euler-Lagrange equation
\begin{equation}
	\zeta_{j \ell}\frac{\md^2\lambda_{\ell}}{\md t^2}+ \frac{\partial\zeta_{j \ell}}{\partial \lambda_{m}} \frac{\md\lambda_{\ell}}{\md t}\frac{\md\lambda_{m}}{\md t} = \frac{1}{2}\frac{\partial\zeta_{\ell m}}{\partial \lambda_{j}} \frac{\md\lambda_{\ell}}{\md t}\frac{\md\lambda_{m}}{\md t} \ ,
	\label{Euler-Lagrange}
\end{equation}
where we have adopted the Einstein convention of implied summation over all repeated indices. We directly calculate the friction matrix from \eqref{CDF friction} and find geodesics by numerically solving \eqref{Euler-Lagrange} with specified initial and final control parameters, as described in Refs.~\onlinecite{Rotskoff2017,louwerse2022}.

{\it Rapidly driven systems.}---In the fast limit, the excess work approaches that of an instantaneous protocol, which spends no time relaxing towards equilibrium and requires excess work proportional (up to a factor of $k_{\rm B}T$) to the \emph{relative entropy} $D(\pi_{\rm i}||\pi_{\rm f})$ between the initial and final equilibrium distributions~\cite{Blaber2021}. Spending a short duration $\Delta t$ relaxing towards equilibrium throughout the protocol results in \emph{saved work} $W_{\rm save} \equiv k_{\rm B}T D(\pi_{\rm i}||\pi_{\rm f}) -W_{\rm ex}$ compared to an instantaneous protocol, which can be approximated as~\cite{Blaber2021}
\begin{align}
\langle W_{\rm save} \rangle_{\boldsymbol{\Lambda}} \approx  \int_{0}^{\Delta t}\md t  \, {\bf R}_{\boldsymbol{\lambda}_{\rm i}}^{\rm T}[\boldsymbol{\lambda}(t)] \, 
[\boldsymbol{\lambda}_{{\rm f}} - \boldsymbol{\lambda}(t)] \ ,
\label{Excess work approx}
\end{align}
in terms of the \emph{initial force-relaxation rate} (IFRR)
\begin{subequations}
	\begin{align}
	{\bf R}_{\boldsymbol{\lambda}_{\rm i}}[\boldsymbol{\lambda}(t)] &\equiv \frac{\md\langle {\bf f}_{\boldsymbol{\lambda}} \rangle_{\boldsymbol{\lambda}_{\rm i}}}{\md t}\bigg|_{\boldsymbol{\lambda}(t)} \ ,
	\label{Rate Function}
	\end{align}
\end{subequations}
the rate of change of the initial mean conjugate forces at the current control-parameter values. 

The saved work is maximized by the \emph{short-time efficient protocol} (STEP) which spends the entire duration at the control-parameter value that maximizes the short-time power savings
\begin{align}
P_{\rm save}^{\rm st}(\boldsymbol{\lambda}) \equiv {\bf R}_{\boldsymbol{\lambda}_{\rm i}}^{\rm T}(\boldsymbol{\lambda})(\boldsymbol{\lambda}_{{\rm f}} - \boldsymbol{\lambda}) \ ,
\label{eq:power_savings}
\end{align}
satisfying
\begin{align}
&\frac{\partial P_{\rm save}^{\rm st}(\boldsymbol{\lambda})}{\partial \boldsymbol{\lambda}}\bigg|_{\boldsymbol{\lambda}^{\rm STEP}}  = 0 \ .
\label{STEP}
\end{align}
The STEP achieves this by two instantaneous control-parameter jumps: one at the start from the initial value to the optimal value $\boldsymbol{\lambda}^{\rm STEP}$, and one at the end from $\boldsymbol{\lambda}^{\rm STEP}$ to the final value. 

For overdamped dynamics, the short-time power saving from the STEP are conveniently expressed as (Supplementary Material~I (SM))
\begin{align}
\label{eq:power_savings_overdamped}
        P_{\rm save}^{\rm st}&(\boldsymbol{\lambda}) = \\ &\frac{1}{\gamma}\left\langle [f_{x}(x,\boldsymbol{\lambda}_{\rm f})-f_{x}(x,\boldsymbol{\lambda})][f_{x}(x,\boldsymbol{\lambda})-f_{x}(x,\boldsymbol{\lambda}_{\rm i})]\right\rangle_{\boldsymbol{\lambda}_{\rm i}} \ , \nonumber
\end{align}
which is maximized if
\begin{align}
\label{STEP_overdamped}
    \left\langle \frac{\partial f_{x}(x,\boldsymbol{\lambda})}{\partial\boldsymbol{\lambda}}\left[f_{x}(x,\boldsymbol{\lambda})-\frac{f_{x}(x,\boldsymbol{\lambda}_{\rm i})+f_{x}(x,\boldsymbol{\lambda}_{\rm f})}{2}\right]\right\rangle_{\boldsymbol{\lambda}_{\rm i}} = 0 \ .
\end{align}
This is achieved by control-parameter values which for all $x$ satisfy $\partial f_{x}(x,\boldsymbol{\lambda})/\partial\boldsymbol{\lambda} = \boldsymbol{0}$ or $f_{x}(x,\boldsymbol{\lambda}) = [f_{x}(x,\boldsymbol{\lambda}_{\rm i})+f_{x}(x,\boldsymbol{\lambda}_{\rm f})]/2$. In what follows we will enforce the second condition, although it is more stringent than \eqref{STEP_overdamped} which only constrains a single average over the entire system distribution.

{\it Model system.}---Here we consider a model system relevant to DNA-hairpin experiments: a Brownian bead driven by a time-dependent quadratic trapping potential with center and stiffness modulated by the focus and intensity of the laser. This model is also typical of steered molecular-dynamics simulations, which use a time-dependent quadratic potential to drive  reactions~\cite{Dellago2014}. The total potential $V_{\rm tot} = V_{\rm hp} + V_{\rm trap}$ is the sum of the static hairpin potential and time-dependent trap potential (shown schematically in Fig.~\ref{Protocol}). The hairpin potential is modeled as a static double well (symmetric for simplicity) with the two minima at $x = 0$ and $x = \Delta x_{\rm m}$ representing the folded and unfolded states~\cite{neupane2015,neupane2017,woodside2006,Sivak2016},
\begin{align}
V_{\rm hp}(x) = E_{\rm B} \left[\left(\frac{2x-\Delta x_{\rm m}}{\Delta x_{\rm m}}\right)^2-1\right]^{2} \ ,
\label{double well}
\end{align}
for barrier height $E_{\rm B}$, distance $x_{\rm m}$ from the minimum to barrier, and distance 
$\Delta x_{\rm m}= 2x_{\rm m}$ between the minima. The system is driven by a quadratic trap
\begin{align}
V_{\rm trap}[x,x^{\rm c}(t),k(t)]= \frac{k(t)}{2}\left[x^{\rm c}(t)-x\right]^2 \ ,
\label{trap potential}
\end{align}
with time-dependent stiffness $k(t)$ and center $x^{\rm c}(t)$.

The total work can be separated into two components, $W = W_{\rm c} + W_{\rm s}$, one for each control parameter. The trap-center component 
\begin{align}
	\langle W_{\rm c} \rangle_{\boldsymbol{\Lambda}} = \int_{0}^{\Delta t} \md t \ k(t) \left[ x^{\rm c}(t)-\langle x\rangle_{\boldsymbol{\Lambda}}\right]\frac{\md x^{\rm c}(t)}{\md t}
\end{align}
is analogous to `force-distance' work. The stiffness component 
\begin{align}
\langle W_{\rm s} \rangle_{\boldsymbol{\Lambda}} = \frac{1}{2}\int_{0}^{\Delta t} \md t  \left\langle \left[x^{\rm c}(t)-x\right]^2\right\rangle_{\boldsymbol{\Lambda}} \frac{\md k(t)}{\md t}
\end{align}
resembles `pressure-volume' work, i.e., the stiffness controls the effective volume available to the system, and the variance contributes to an effective pressure resisting changes in trap stiffness.

{\it Designed protocols.}---We consider protocols that drive the system between the two minima $x^{\rm c}_{\rm i} = 0$ and $x^{\rm c}_{\rm f} = \Delta x_{\rm m}$ with equal initial and final stiffness ($k_{\rm i} = k_{\rm f}$). This allows us to directly compare control over only trap center to control over both trap center and stiffness. We speculate that for unequal initial and final stiffness the qualitative features of the designed protocols would remain similar.

In the slow limit, the two-dimensional linear-response (2D LR) protocol minimizes dissipation by tightening the trap and slowing down as it traverses the barrier (Fig.~\ref{Work_distance}). Tightening the trap when approaching the barrier (Fig.~\ref{Work_distance}b) helps the system maintain roughly constant variance throughout the protocol and approximately linearly changes the quantiles of the position distribution (Fig.~1), which is a generic property for minimum-dissipation protocols in optimal transport under full control~[39-41]. When the trap doesn't tighten (e.g., naive and 1D LR protocols in Fig.~2b), the variance increases as the system crosses the barrier and the quantiles do not change linearly. Slowing down while crossing the barrier (previously observed for one-dimensional (constant-stiffness) barrier crossing~[34]) allows time for thermal fluctuations to kick the system over the barrier (Fig.~\ref{Work_distance}a).

In SM~II we show that the amount that the trap tightens and slows down depends on the initial stiffness. We call the initial stiffness large (small) when the initial energy of the trap at the barrier $x_{\rm m}$ is significantly larger (smaller) than $E_{\rm b}$; i.e., $k_{\rm i} \gg E_{\rm b}/x_{\rm m}^2$ ($k_{\rm i} \ll E_{\rm b}/x_{\rm m}^2$). Physically, large (small) initial stiffness ensures that the initial equilibrium distribution is unimodal (bimodal). Throughout, we compare the stiffness to the \emph{scaled barrier height} $E_{\rm B}/x_{\rm m}^2$, essentially comparing the initial energy of the trap and hairpin potential at the barrier. The 2D LR protocol changes stiffness most when the initial stiffness is comparable to the scaled barrier height ($k_{\rm i} \sim E_{\rm B}/x_{\rm m}^2$), and leaves stiffness virtually unchanged when the initial stiffness is either large ($k_{\rm i} \gg E_{\rm B}/x_{\rm m}^2$) or small ($k_{\rm i} \ll E_{\rm B}/x_{\rm m}^2$). 

The minimum-dissipation protocol in the fast limit (the STEP) maximizes the short-time power saving~\eqref{eq:power_savings_overdamped} by jumping from and to the control-parameter endpoints to spend the entire duration at control-parameter values $x^{\rm STEP} = (x^{\rm c}_{\rm i}+x^{\rm c}_{\rm f})/2$, and $k^{\rm STEP} =  k_{\rm i}$. This result is independent of the hairpin potential since $f_{x}(x,x^{\rm STEP},k^{\rm STEP}) = [f_{x}(x,x^{\rm c}_{\rm i},k_{\rm i})+f_{x}(x,x^{\rm c}_{\rm f},k_{\rm i})]/2$ maximizes the short-time power saving \eqref{STEP_overdamped} for all $x$ independent of $V(x)$ (note $k_{\rm f} = k_{\rm i}$).

Given theory describing minimum-dissipation control in both the slow and fast limits, we develop a simple interpolation scheme to design protocols that reduce dissipation at all driving speeds. Similar to the one-dimensional case~\cite{Blaber2021}, we choose the interpolated protocol to have an initial jump $(\lambda^{\rm STEP}-\lambda_{\rm i})/(1+\Delta t/\tau)$ and a final jump $(\lambda_{\rm f}-\lambda^{\rm STEP})/(1+\Delta t/\tau)$, and follow the original linear-response path between them, 
\begin{align}
    \lambda^{\rm interp}(t) = \frac{1}{1+\frac{\Delta t}{\tau}}\lambda^{\rm STEP} + \left(1-\frac{1}{1+\frac{\Delta t}{\tau}}\right)\lambda^{\rm LR}(t)\ ,
\end{align}
with $\tau$ the crossover duration. This guarantees that the protocol approaches the minimum-dissipation protocol in both the fast and slow limits. For system timescale, we choose (primarily for its simplicity) the diffusion time $\tau_{\rm D} \equiv \Delta x_{\rm m}^2/(2D)$ between wells, for $D \equiv (\beta\gamma)^{-1}$. More sophisticated measures of relaxation time~\cite{Li2019,louwerse2022} could yield improved performance of the interpolated protocol.

{\it Performance.}---For comparison to an ideal process, we evaluate the performance of optimal-transport (OT) theory under full control~\cite{Aurell2011,abreu2011,Proesmans2020}. We use optimal transport to calculate the minimum work required---assuming complete control over the potential---to move probability from an initial to final distribution within a fixed duration as
\begin{align}
    W_{\rm OT} = F_{\rm f} - F_{\rm i} + \frac{\beta\gamma}{\Delta t}\int_{0}^{1}\md y \left[Q_{\rm f}(y)-Q_{\rm i}(y)\right]^2 \ ,
\end{align}
where $Q_{\rm f}$ and $Q_{\rm i}$ are the final and initial quantile functions (inverse cumulative distribution functions). We then perform a second optimization over the final probability distribution, subject to constrained initial and final control-parameter endpoints. This yields the minimum work to drive the trap between the two endpoints if we had full control over the potential (rather than just one- or two-dimensional parametric control).

Figure~\ref{Work_distance} compares naive, one-dimensional linear-response (1D LR), two-dimensional linear-response (2D LR), and interpolated protocols. We measure performance by the average work (the direct target of the protocol design) and the probability $p_{\boldsymbol{\Lambda}}(x<x_{\rm m})$ that the system remains in its initial well (related to the average flux) which was not directly considered in the design. The barrier height $E_{\rm B} = 4k_{\rm B} T$ is intermediate in the context of DNA hairpins~\cite{woodside2006,neupane2017}, and the initial and final stiffness are comparable to the scaled barrier height, $k_{\rm i} = k_{\rm f} = E_{\rm B}/x_{\rm m}^2$. For a $1$-$\mu$m bead in water at standard temperature and pressure, with inter-well distance  $\Delta x_{\rm m} \approx 20$ nm, the initial and final stiffness correspond to $k_{\rm i} = k_{\rm f} \approx 0.16$ pN/nm, the 2D LR protocol reaches a maximum stiffness of $k_{\rm max} \approx 1.12$ pN/nm, and the diffusion time between wells is $\tau_{\rm D} \approx 0.4$~s.

For long duration ($\Delta t \gg \tau_{\rm D}$), the 1D LR protocol requires ($\sim$1.6$\times$) less work than the naive; however, the 2D LR and interpolated protocols most significantly reduce work (Fig.~\ref{Work_distance}c; $\sim$5.6$\times$ less than naive, $\sim$3.5$\times$ less than 1D LR, and within 1\% of full control). Intermediate-duration designed protocols give the largest-magnitude work reduction $\langle W \rangle_{\rm des} - \langle W \rangle_{\rm naive}$: 2D LR and interpolated protocols save $\sim$2.7$\,k_{\rm B} T$, whereas 1D LR only saves $\sim$0.4$\,k_{\rm B} T$ (Fig.~\ref{Work_distance}e).

1D LR protocols often reduce dissipation but as a side effect also decrease flux, as seen in Fig.~\ref{Work_distance}d. 2D LR and interpolated protocols have the opposite effect, decreasing dissipation while increasing flux. For intermediate duration, the 2D LR protocol drives up to 78\% and the interpolated up to 17\% more probability to the destination well, compared to naive; the 1D LR drives 19\% \emph{less}. 

For long duration ($\Delta t \gg \tau_{\rm D}$), two-dimensional control provides significant advantages over one-dimensional control for both average work and flux; however, for short duration the 2D LR protocol can perform worse than 1D LR and naive (Fig.~\ref{Work_distance}c and e; similar behavior has been observed for multidimensional control of the Ising model~\cite{louwerse2022}). For short duration, the system cannot keep up with the rapid changes in the trap potential, and the linear-response approximation breaks down.  Although the increased stiffness of the 2D LR protocol results in the strongest driving and hence the greatest flux of the protocols considered here (Fig.~2), it does so at the cost of increased dissipation for short duration. Indeed, the minimum-dissipation protocol for short duration ($\Delta t \ll \tau_{\rm D}$, the STEP) is monotonic and discrete. Our interpolated protocol asymptotes to the STEP in the short-duration limit, resulting in reduced dissipation and increased flux at any duration. In terms of dissipation, the interpolated protocol achieves within 1\% of the minimum work under full control for short and long duration and remains within 30\% of full control at intermediate duration (Fig.~\ref{Work_distance}c).

{\it Discussion.}---We have shown that multidimensional control protocols can significantly outperform their one-dimensional counterparts, improving both work and flux. For a system undergoing driven barrier crossing, one-dimensional control of only the trap center limits the control over the position distribution, with a large increase in variance as the protocol crosses the barrier (Fig.~\ref{Protocol}). Control over both the trap center and stiffness makes possible approximately linear driving of the position mean and variance between specified endpoints, consistent with optimal-transport protocols that minimize work under full control~\cite{Aurell2011,abreu2011,Proesmans2020}. This significantly reduces the work required to drive the system between the two wells and increases the flux compared to naive and one-dimensional control protocols (Fig.~\ref{Work_distance}). The main shortcoming of the multidimensional linear-response protocols is that they can perform worse than naive for short duration; however, we remedy this issue by combining linear-response and STEP frameworks to give interpolated protocols that reduce dissipation at any duration. For the model system and parameters we explored, the largest reduction in dissipation occurs from one- to two-dimensional control, and the dissipation in the two-dimensional interpolated protocol is within 30\% of full control for intermediate duration and within 1\% for short and long duration.

The model system closely resembles DNA-hairpin experiments, and we explore experimentally relevant parameters~\cite{woodside2006,neupane2017}. Our results reveal general design principles for driven barrier crossing that can be readily implemented experimentally: the designed protocols 1) slow down and tighten the trap as it crosses the energy barrier, thereby driving the mean position between the two wells at constant rate while maintaining constant variance; and 2) jump at the beginning and end of the protocol, with larger jumps for faster protocols. Recent experimental protocols implement one-dimensional control and demonstrate significant work reductions from designed protocols~\cite{Tafoya2019}. We show that adding an additional control parameter (trap stiffness) can dramatically improve the performance over the one-dimensional counterpart (up to $3.5\times$ less work and ~80\% increased probability of reaching the target well). Although multidimensional control is more difficult to implement, the performance gains can be significant.

\section*{Acknowledgement}
This work is supported by an SFU Graduate Deans Entrance Scholarship (SB), an NSERC Discovery Grant and Discovery Accelerator Supplement (DAS), and a Tier-II Canada Research Chair (DAS), and was enabled in part by support provided by WestGrid (www.westgrid.ca) and Compute Canada Calcul Canada (www.computecanada.ca). The authors thank Miranda Louwerse (SFU Chemistry) for sharing code used to determine geodesics of the friction metric and enlightening feedback on the manuscript.

\onecolumngrid
\clearpage
\begin{center}
	\textbf{\large Supplemental Material for ``Efficient two-dimensional control of barrier crossing''}
\end{center}
\setcounter{equation}{0}
\setcounter{figure}{0}
\setcounter{table}{0}
\setcounter{page}{1}
\makeatletter
\renewcommand{\theequation}{S\arabic{equation}}
\renewcommand{\thefigure}{S\arabic{figure}}

\section{\label{App Fast Limit} Fast Limit}
In this section we derive a simple formula for the short-time power savings used to determine the STEP in the \emph{Rapidly driven systems} section of the main text.

For a rapidly driven system, the minimum-dissipation protocol consists of two discrete jumps, spending the entire duration at fixed control-parameter values. This type of discrete protocol requires work
\begin{align}
    \langle W \rangle_{\boldsymbol{\Lambda}} = \big\langle V_{\rm tot}(x,\boldsymbol{\lambda}_{\rm f}) - V_{\rm tot}(x,\boldsymbol{\lambda}) \big\rangle_{\boldsymbol{\Lambda}}+\big\langle V_{\rm tot}(x,\boldsymbol{\lambda}) - V_{\rm tot}(x,\boldsymbol{\lambda}_{\rm i}) \big\rangle_{\boldsymbol{\lambda}_{\rm i}} \ .
    \label{Discrete work}
\end{align}
For short duration $\Delta t$, we approximate the probability distribution at the conclusion of the protocol as~\cite{Blaber2021}
\begin{subequations}
\begin{align}
    p_{\boldsymbol{\Lambda}}(x,\Delta t) &\approx \pi_{\rm i}(x) + \Delta t \,  L(x,\boldsymbol{\lambda})\pi_{\rm i}(x) \\
    &\approx \pi_{\rm i}(x) + \Delta t \,  \left[ L(x,\boldsymbol{\lambda}) - L(x,\boldsymbol{\lambda}_{\rm i})\right]\pi_{\rm i}(x) \ ,
\end{align}
\end{subequations}
where $L(x,\boldsymbol{\lambda})$ is the time-evolution operator for the probability distribution at fixed control parameter $\boldsymbol{\lambda}$, and in the second line we have used the fact that the initial equilibrium distribution satisfies $L(x,\boldsymbol{\lambda}_{\rm i})\pi_{\rm i}(x) = 0$. For Fokker-Planck dynamics this gives
\begin{align}
        p_{\boldsymbol{\Lambda}}(x,\Delta t) \approx \pi_{\rm i}(x) + \frac{\Delta t}{\gamma} \frac{\partial }{\partial x}\Big\{\big[f_{x}(x,\boldsymbol{\lambda}) - f_{x}(x,\boldsymbol{\lambda}_{\rm i})\big]\pi_{\rm i}(x)\Big\} \ .
\end{align}
Substituting into \eqref{Discrete work} and rearranging gives
\begin{subequations}
\begin{align}
    \langle W_{\rm save} \rangle_{\boldsymbol{\Lambda}} &\equiv \big\langle V_{\rm tot}(x,\boldsymbol{\lambda}_{\rm f})-V_{\rm tot}(x,\boldsymbol{\lambda}_{\rm i})\big\rangle_{\boldsymbol{\lambda}_{\rm i}} - \langle W \rangle_{\boldsymbol{\Lambda}} \\
    &= \frac{\Delta t}{\gamma}\int\md x \, \big[V_{\rm tot}(\boldsymbol{\lambda})-V_{\rm tot}(\boldsymbol{\lambda}_{\rm f})\big] \, \frac{\partial }{\partial x}\Big\{\big[f_{x}(x,\boldsymbol{\lambda}) - f_{x}(x,\boldsymbol{\lambda}_{\rm i})\big] \, \pi_{\rm i}(x)\Big\} \ ,
\end{align}
\end{subequations}
Integrating by parts leads to
\begin{align}
    \langle W_{\rm save} \rangle_{\boldsymbol{\Lambda}} = \frac{\Delta t}{\gamma}\Big\langle \big[f_{x}(x,\boldsymbol{\lambda}_{\rm f})-f_{x}(x,\boldsymbol{\lambda})\big]\, \big[f_{x}(x,\boldsymbol{\lambda}) - f_{x}(x,\boldsymbol{\lambda}_{\rm i})\big]\Big\rangle_{\boldsymbol{\lambda}_{\rm i}} \ .
\end{align}
The short-time power saving is $P_{\rm save}^{\rm st}(\boldsymbol{\lambda}) = \langle W_{\rm save} \rangle_{\boldsymbol{\Lambda}}/\Delta t$, which can be expressed as
\begin{align}
    P_{\rm save}^{\rm st}(\boldsymbol{\lambda}) = \frac{1}{\gamma}\Big\langle \big[f_{x}(x,\boldsymbol{\lambda}_{\rm f})-f_{x}(x,\boldsymbol{\lambda})\big]\big[f_{x}(x,\boldsymbol{\lambda})-f_{x}(x,\boldsymbol{\lambda}_{\rm i})\big]\Big\rangle_{\boldsymbol{\lambda}_{\rm i}} \ ,
\end{align}
and is maximized if
\begin{align}
    \left\langle \frac{\partial f_{x}(x,\boldsymbol{\lambda})}{\partial\boldsymbol{\lambda}}\left[f_{x}(x,\boldsymbol{\lambda})-\frac{f_{x}(x,\boldsymbol{\lambda}_{\rm i})+f_{x}(x,\boldsymbol{\lambda}_{\rm f})}{2}\right]\right\rangle_{\boldsymbol{\lambda}_{\rm i}} = 0 \ .
\end{align}
This can be achieved by control parameters which for all $x$ satisfy $\partial f_{x}(x,\boldsymbol{\lambda})/\partial\boldsymbol{\lambda} = 0$ or $f_{x}(x,\boldsymbol{\lambda}) = [f_{x}(x,\boldsymbol{\lambda}_{\rm i})+f_{x}(x,\boldsymbol{\lambda}_{\rm f})]/2$.

\section{\label{App Slow Limit} Slow Limit}
In this section we describe in detail the designed protocols (geodesics) based on the friction matrix for driven barrier crossing under control of both trap center and stiffness. We demonstrate that the largest change in stiffness from the designed protocols occurs when the stiffness is comparable to the scaled barrier height ($k_{\rm i} \sim E_{\rm b}/x_{\rm m}^2$), as discussed in the \emph{Designed protocols} section of the main text.

\begin{figure}
	\includegraphics[width=0.9\linewidth]{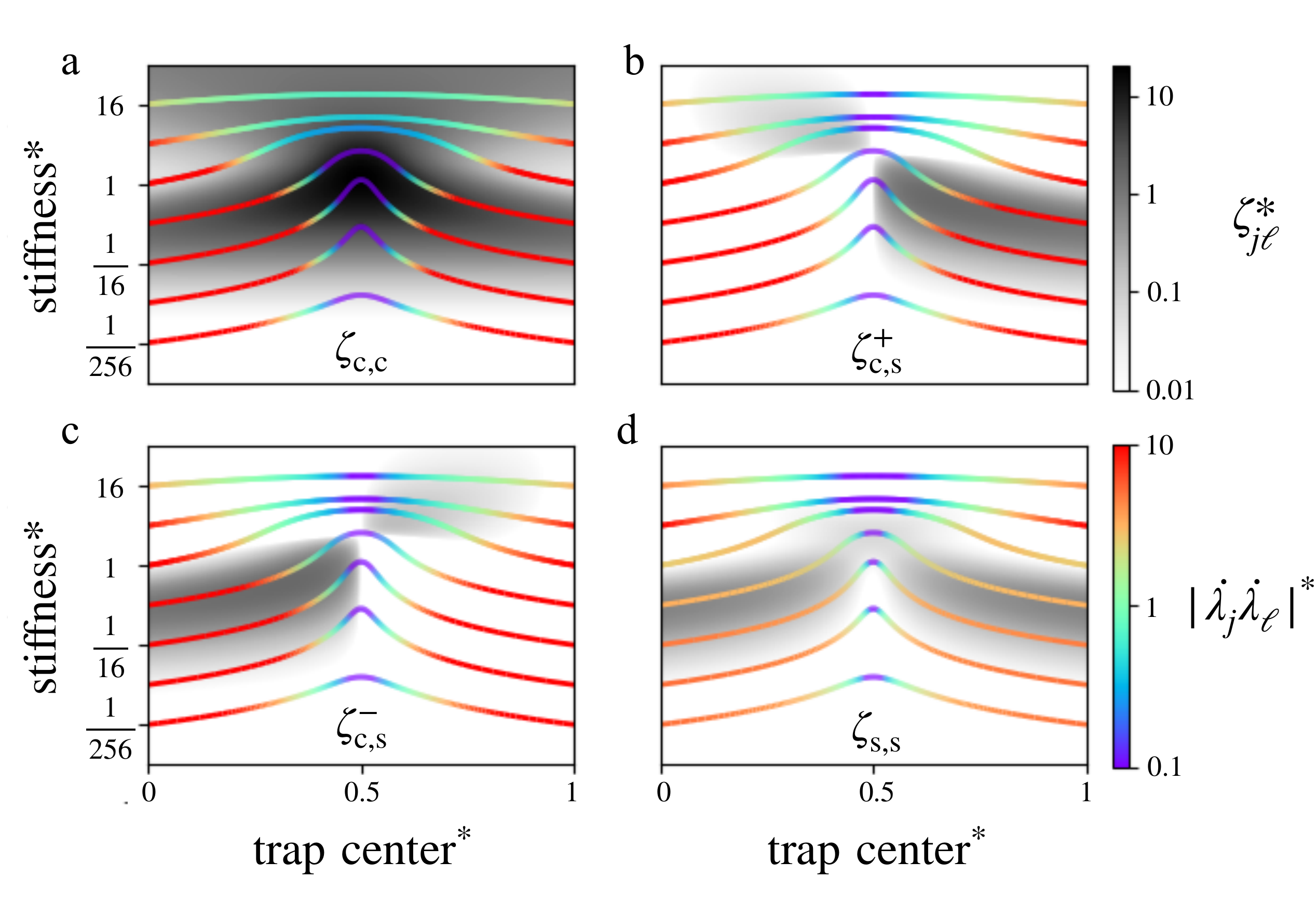} 
	\caption{Geodesics and components of the friction matrix used to design two-dimensional linear-response protocols. Grayscale heatmap: components of the friction as a function of the (dimensionless) trap center$^*$ $x^{\rm c}/{\Delta x_{\rm m}}$ and stiffness$^*$ $k x_{\rm m}^{2}/E_{\rm B}$. Colored curves: geodesics of the friction for equal initial and final trap stiffness ($k_{\rm i} = k_{\rm f}$). Color heatmap: absolute product of control-parameter speeds $\dot{\lambda}_{j} = \md \lambda_{j} / \md t$. The positive and negative components of the off-diagonal entry $\zeta_{\rm c,s}$ are respectively denoted by $\zeta_{\rm c,s}^{+}$ (b) and $\zeta_{\rm c,s}^{-}$ (c). A star denotes a scaled (dimensionless) quantity, with the velocities scaled by the average speed $|\dot{\lambda}_{j}\dot{\lambda}_{\ell}|^* \equiv |\dot{\lambda}_{j}\dot{\lambda}_{\ell}|/(\overline{|\dot{\lambda}_{j}|}~\overline{|\dot{\lambda}_{\ell}|})$ and friction as $\zeta_{j\ell}^* \equiv \zeta_{j\ell}\lambda_{j}\lambda_{\ell}/(\lambda_{j}^{*}\lambda_{\ell}^{*}\gamma x_{\rm m}^2)$.}
	\label{Geodesics}
\end{figure}

For slow driving, the excess work is described by \eqref{Excess work approx}, where the friction can be directly calculated from \eqref{CDF friction} to yield the friction matrix shown in Fig.~\ref{Geodesics}. The geodesics are found by numerically solving \eqref{Euler-Lagrange} with specified initial and final trap center and stiffness, as described in Refs.~\onlinecite{Rotskoff2017,louwerse2022}. We consider protocols that drive the system between the two minima $x^{\rm c}_{\rm i} = 0$ and $x^{\rm c}_{\rm f} = \Delta x_{\rm m}$ with equal initial and final stiffness, $k_{\rm i} = k_{\rm f}$. 

All components of the friction have the largest variation in magnitude across the protocol when the trap stiffness is comparable to the scaled barrier height ($k\sim E_{\rm b}/x_{\rm m}^2$). If the stiffness is small ($k\ll E_{\rm b}/x_{\rm m}^2$) or large ($k\gg E_{\rm b}/x_{\rm m}^2$), then all components of the friction are independent of the trap center. For $k\ll E_{\rm b}/x_{\rm m}^2$, the total potential~\eqref{Fokker Planck} is dominated by the hairpin potential (independent of the trap potential), so the friction~\eqref{friction} is independent of the trap center and stiffness. For $k\gg E_{\rm b}/x_{\rm m}^2$, the total potential is dominated by the trap potential (independent of the hairpin potential), and the friction approaches that of a harmonic trap on a flat landscape, which is also independent of the trap center.

The center-center component $\zeta_{\rm c,c}$ of the friction matrix is strongly peaked at the barrier (Fig.~\ref{Geodesics}). This component is proportional to the force variance $\langle(\delta f_{x^{\rm c}})^2\rangle_{\boldsymbol{\lambda}} = k\langle (\delta x)^2 \rangle_{\boldsymbol{\lambda}}$, which is largest in magnitude at the barrier. The barrier reduces the effective stiffness of the total potential, thereby increasing the position variance. Physically, a distribution sharply peaked at the trap center requires less work to translate than a wider distribution.

The stiffness-stiffness component $\zeta_{\rm s,s}$ of the friction is proportional to the fourth moment of the position distribution, $\langle(\delta f_{k})^2\rangle_{\boldsymbol{\lambda}} = \langle [\delta (x-x^{\rm c})^2]^2 \rangle_{\boldsymbol{\lambda}}/4$, which is largest when the distribution has appreciable probability of extreme values. Therefore, this component of the friction is largest when the total potential is a double well with two widely separated wells. For $k \lesssim E_{\rm B}/x_{\rm m}^2$, there is significant probability in the well opposite the trap (i.e., the distribution is bimodal), and pulling the trap closer to the center reduces the friction by reducing the distance between the two minima of the total potential. For $k \gtrsim E_{\rm B}/x_{\rm m}^2$, the total potential only has one minimum, and therefore this component of the friction is largest when the (unimodal) position distribution is widest, which occurs at the barrier. Physically, it takes more work to tighten the trap when the system is far from the trap center, scaling as $(x-x^{\rm c})^4$.

The off-diagonal component $\zeta_{\rm c,s}$ has both positive contributions, $\zeta_{\rm c,s}^{+} \equiv \max(\zeta_{\rm c,s},0)$, and negative contributions, $\zeta_{\rm c,s}^{-}\equiv \max(-\zeta_{\rm c,s},0)$. The off-diagonal components result from cross-correlations between the conjugate forces, and can either increase or decrease the work compared to treating the conjugate forces as uncorrelated (ignoring off-diagonal components).

First we consider a weak trap, $k \lesssim E_{\rm B}/x^{2}_{\rm m}$. For $x^{\rm c} < \Delta x_{\rm m}/2$, $\zeta_{\rm c,s}$ is negative so increasing or decreasing both the trap center and stiffness together results in negative contribution to the total work from this component. Tightening the trap as the system is driven over the barrier causes this contribution to reduce the total work. For $x^{\rm c} > \Delta x_{\rm m}/2$, $\zeta_{\rm c,s}$ is positive so the contribution to the total work is negative if the trap center is increased as the stiffness is decreased. Loosening the trap as it is driven away from the barrier causes this contribution to decrease the total work.

For a strong trap ($k \gtrsim E_{\rm B}/x^{2}_{\rm m}$), the situation is reversed: tightening the trap as the system is driven up the energy landscape and loosening the trap as it is driven down result in a positive total-work contribution from the off-diagonal component. Since the trap is stiff compared to the hairpin potential, tightening no longer helps pull the system up the energy landscape and instead tightly confines the system, attenuating thermal fluctuations which would otherwise help kick the system over the barrier.

\end{document}